\documentclass[12pt,a4paper]{article}
\pdfoutput=1

\usepackage{a4wide}
\usepackage{epsfig}
\usepackage{graphicx}
\usepackage{multicol}
\usepackage{slashed}
\usepackage[intlimits]{amsmath}
\usepackage{amssymb}
\usepackage[small]{caption}
\usepackage{cite}
\usepackage[usenames,dvipsnames]{color}
     \definecolor{hgreen}{rgb}{0,.3,0}
     \definecolor{hred}{rgb}{.3,0,0}
     \definecolor{hblue}{rgb}{0,0,.3}
     \definecolor{LightGray}{gray}{0.95}
\usepackage[
            colorlinks=true,
            linkcolor=hblue,
            citecolor=hgreen,
            filecolor=hblue,
            urlcolor=hred]{hyperref}

\numberwithin{equation}{section}
\newcommand{\GF}{G_{\rm F}}

\newcommand{\Bsmumu}{B_s\to\mu^+\mu^-}
\newcommand{\Bdmumu}{B_d\to\mu^+\mu^-}
\newcommand{\Bqmumu}{B_q\to\mu^+\mu^-}

\providecommand{\doi}[1]{\href{http://dx.doi.org/#1}{doi:#1}}

\newcommand{\ttZ}{t\bar t Z}
\newcommand{\bbZ}{b\bar b Z}

\begin{document}


\title{ \boldmath \textbf{Probing anomalous $\ttZ$ interactions with 
rare meson decays} \unboldmath }

\author{
{
$\text{Joachim Brod}^{a}$\footnote{joachim.brod@uc.edu}, 
$\text{Admir Greljo}^{b}$\footnote{admir.greljo@ijs.si}, 
$\text{Emmanuel Stamou}^{c}$\footnote{emmanuel.stamou@weizmann.ac.il}, 
$\text{Patipan Uttayarat}^{a,d}$\footnote{patipan@g.swu.ac.th}}\\[2em]
  {\normalsize $^{a}$Department of Physics, University of Cincinnati, Cincinnati, Ohio 45221, USA}\\[0.5em] 
  {\normalsize $^{b}$Jo\v{z}ef Stefan Institute, Jamova 39, 1000 Ljubljana, Slovenia}\\[0.5em] 
  {\normalsize $^{c}$Department of Particle Physics and Astrophysics, Weizmann Institute of Science,}\\
  {\normalsize Rehovot 76100, Israel}\\[0.5em]
  {\normalsize $^{d}$Department of Physics, Srinakharinwirot University, Bangkok 10110, Thailand}\\[0.5em]
}

\date{}

\maketitle

\begin{abstract}
\addcontentsline{toc}{section}{Abstract}
Anomalous couplings of the $Z$ boson to top quarks are only marginally
constrained by direct searches and are still sensitive to new particle
dynamics at the TeV scale. Employing an effective field theory
approach we consider the dimension-six operators which generate
deviations from the standard-model vector and axial-vector
interactions. We show that rare $B$ and $K$ meson decays together with
electroweak precision observables provide strong constraints on these
couplings. We also consider constraints from t-channel single-top
production.

\end{abstract}


\section{Introduction
\label{sec:introduction}}

In the case of no direct  observation of 
new particles at the Large Hadron Collider (LHC), precision measurements of
the Standard Model (SM) interactions provide an alternative route to
discover New Physics (NP).
The measurement of the $\ttZ$ production cross section at the
LHC~\cite{ATLAS-CONF-2012-126,Chatrchyan:2013qca} offers a direct test
of anomalous $\ttZ$ couplings. In a recent publication, Schulze and
R\"ontsch presented the calculation and analysis of $\ttZ$ production
at next-to-leading order in QCD~\cite{Rontsch:2014cca}.  They used
current and future-projected LHC data to constrain anomalous vector
and axial-vector couplings of the $Z$ boson to top quarks.

In this work we study constraints on anomalous $\ttZ$ couplings
arising from electroweak precision observables (EWPO), rare $B$ and
$K$ meson decays, and t-channel single-top production.  The $\ttZ$
couplings contribute to the rate of rare meson decays and to 
EWPO via radiative corrections. Thus, in the absence of contributions to these
observables from other sources, rare meson decays and EWPO offer
strong bounds on anomalous $\ttZ$ couplings. In fact, we show that the
indirect constraints exceed the direct bounds, even after the
projected high-luminosity upgrade of the LHC~\cite{Rontsch:2014cca}.

We focus on the rare decays $B_s \to \mu^+ \mu^-$, $K^+ \to \pi^+ \nu
\bar\nu$ and $K_L \to \pi^0 \nu \bar \nu$. These decays proceed via
$Z$-penguin diagrams\footnote{The photon penguin does not contribute
  to $B_s \to \mu^+ \mu^-$ due to conservation of the vector current.}
which, in the SM, are dominated by the top-quark loop.  This
makes these processes especially sensitive to anomalous $\ttZ$
couplings. For this reason, the current observation of the $B_s \to \mu^+ \mu^-$
rate already competes with constraints from EWPO and the expected future 
improvements in the measurements of rare meson decay rates are 
especially encouraging as they may provide the first evidence for 
anomalous $\ttZ$ interactions.

The use of rare decays to obtain bounds on anomalous $Z$ couplings has
already been discussed in
Refs.~\cite{Haisch:2007ia,Guadagnoli:2013mru}.  The authors derived
constraints on $b\bar bZ$ couplings by relating them to
flavor-changing couplings within the minimal flavor violation (MFV)
framework.  In this work, we consider constraints arising from
operator mixing, induced by electroweak loops, within an effective
field theory (EFT) approach.  This strategy has already been proposed
in Refs.~\cite{Li:2011af, Gong:2013sh} for flavor-changing neutral
currents (FCNCs) involving the top quark, as well as for constraining
anomalous $Wtb$ couplings~\cite{Drobnak:2011wj, Drobnak:2011aa}.

The remainder of the paper is structured as follows: In
Sec.~\ref{sec:tbph} we define the appropriate EFT approach to study
possible deviations in $\ttZ$ couplings. In Sec.~\ref{sec:pheno} we
revisit the main constraints from EWPO and t-channel single-top
production, and present our analysis of rare meson decays.  In
Sec.~\ref{sec:discussion} we discuss the numerical results and
conclude.

\section{Effective field theory\label{sec:tbph}}

Our starting point is the SM augmented by gauge-invariant
dimension-six operators constructed out of the SM fields. This may be
thought of as the effective field theory after integrating out NP at
some scale~$\Lambda$ significantly larger than the electroweak scale. We write
the effective Lagrangian as
\begin{equation}\label{eq:lag_unbroken}
\begin{split}
  {\cal L}^\text{eff} = {\cal L}^\text{SM} + \sum_{k}
  \frac{1}{\Lambda^2} C_{k} Q_{k} \,.
\end{split}
\end{equation}
We are interested in the subset of operators that induce modifications
of the vector and axial-vector $\ttZ$ couplings at tree-level. They
will have the main impact on direct searches for anomalous $\ttZ$
couplings. If we assume no correlations between Wilson coefficients, a
complete set of such operators is given by
(cf.~\cite{AguilarSaavedra:2008zc,Grzadkowski:2010es}):
\begin{equation}\label{eq:basis_unbroken}
\begin{split}
  Q_{\phi q,33}^{(3)} & \equiv (\phi^\dagger i\! \stackrel{\leftrightarrow}{D_\mu^a} \phi) (\bar Q_{L,3} \gamma^\mu \sigma^a Q_{L,3}) \,, \\
  Q_{\phi q,33}^{(1)} & \equiv (\phi^\dagger i\! \stackrel{\leftrightarrow}{D}_\mu \phi) (\bar Q_{L,3} \gamma^\mu Q_{L,3}) \,, \\
  Q_{\phi u,33} & \equiv (\phi^\dagger i\! \stackrel{\leftrightarrow}{D}_\mu \phi) (\bar t_{R} \gamma^\mu t_{R}) \,.
\end{split}
\end{equation}
These operators contain the Higgs doublet $\phi$, the left-handed
third-generation quark doublet $Q_{L,3}$, and the right-handed top
quark $t_R$. Moreover, $\sigma^a$ are the Pauli matrices and $D_\mu$
is the SM gauge-covariant derivative and we defined
\begin{equation}
\begin{split}
  (\phi^\dagger i\! \stackrel{\leftrightarrow}{D_\mu} \phi) & = i
  \phi^\dagger \big(D_\mu \phi \big) - i \big(D_\mu \phi \big)^\dagger 
  \phi \,, \\
  (\phi^\dagger i\! \stackrel{\leftrightarrow}{D_\mu^a} \phi) & = i
  \phi^\dagger \sigma^a \big(D_\mu \phi \big) - i \big(D_\mu \phi
  \big)^\dagger \sigma^a \phi \,,
\end{split}
\end{equation}
so that the operators are manifestly Hermitian. Therefore, all Wilson
coefficients considered in this work are real. 

In order to include in Eq.~\eqref{eq:basis_unbroken} all operators that 
induce $\ttZ$ at tree-level, we have chosen the basis in which the up-type 
quark Yukawa matrix is diagonal. Accordingly, we have
\begin{equation}
  Q_{L,3} \equiv \left[\begin{array}{r}t_{L}\\ \sum_j V_{3j}d_{L,j}\end{array}\right]\,.
\end{equation}
The fields $t_{L(R)}$, $d_{L,j}$ are mass eigenstates and $V$ is
the Cabibbo-Kobayashi-Maskawa (CKM) matrix. 
A generic NP model can generate FCNC transitions in the up-quark
sector; to take these effects into account we have to consider
additional operators involving the first and second quark
generation. However, in the restricted, but phenomenologically well-motivated,
framework of MFV~\cite{D'Ambrosio:2002ex} these operators are 
suppressed with respect to those in Eq.~\eqref{eq:basis_unbroken} by
elements of the CKM matrix. Thus, the resulting bounds on $\ttZ$
couplings are negligible. Moreover, in the limit that only the
top-quark Yukawa is non-vanishing in the MFV spurion counting, 
such additional operators are absent. For simplicity, we will use this
approximation in the following, but we will comment on the effect of
keeping a large bottom-quark Yukawa in Sec.~\ref{sec:discussion}. This 
discussion will cover MFV models with down-type quark alignment, i.e. models
in which, at tree-level, only up-type quark FCNCs are generated by NP.

In our analysis, only the operators in Eq.~\eqref{eq:basis_unbroken}
receive non-zero initial conditions at the scale $\Lambda$. However,
electroweak corrections and corrections involving the SM top-quark
Yukawa coupling $y_t$ will induce mixing into the following additional
operators relevant for our analysis:
\begin{equation}\label{eq:basis_unbroken_NLO}
\begin{split}
  Q_{\phi q,ii}^{(3)} & \equiv (\phi^\dagger i\! \stackrel{\leftrightarrow}{D_\mu^a} \phi) (\bar Q_{L,i} \gamma^\mu \sigma^a Q_{L,i}) \,, \\
  Q_{\phi q,ii}^{(1)} & \equiv (\phi^\dagger i\! \stackrel{\leftrightarrow}{D}_\mu \phi) (\bar Q_{L,i} \gamma^\mu Q_{L,i}) \,, \\
  Q_{lq,33jj}^{(3)} & \equiv (\bar Q_{L,3} \gamma_\mu \sigma^a Q_{L,3}) (\bar L_{L,j} \gamma^\mu \sigma^a L_{L,j}) \,, \\
  Q_{lq,33jj}^{(1)} & \equiv (\bar Q_{L,3} \gamma_\mu Q_{L,3}) (\bar L_{L,j} \gamma^\mu L_{L,j}) \,, \\
  Q_{\phi D} & \equiv \big| \phi^\dagger D_\mu \phi \big|^2 \,,
\end{split}
\end{equation}
where $i=1,2$ and $j=1,2,3$. We can safely neglect the remaining
Yukawa couplings in the mixing calculation.  The operators in
Eq.~\eqref{eq:basis_unbroken_NLO} will lead to additional bounds on
the anomalous $\ttZ$ couplings.

To study the effects of these operators on rare decays we
need to know their Wilson coefficients at the electroweak scale where
we match to the five-flavor effective theory. The running of the
Wilson coefficients is given by the renormalization group equations
(RGE). The matching is most conveniently performed in the broken phase
where the Higgs field obtains its vacuum-expectation value, $v$. The relevant part
of the effective Lagrangian in the broken phase reads 
\begin{equation}\label{eq:LSMeffbroken}
\begin{split}
  {\cal L}^\text{eff} \supset {\cal L}^\text{SM} + \frac{v^2}{\Lambda^2}
  \big( {\cal L}^\text{diag}_Z + {\cal L}^\text{FCNC}_Z  +
      {\cal L}^\text{}_W \big) \,.
\end{split}
\end{equation}
Using the conventions of Ref.~\cite{Denner:1991kt}, the
flavor-diagonal neutral-current part of the Lagrangian is given by
\begin{equation}
\begin{split}
  {\cal L}^\text{diag}_Z &= \frac{e}{2 s_w c_w} Z_\mu \Big[
    \big(C_{\phi q,33}^{(3)} - C_{\phi q,33}^{(1)}\big) \, \bar t_L \gamma^\mu t_L -
    C_{\phi u,33} \, \bar t_R \gamma^\mu t_R \\ & \hspace{2.5cm} -
    \sum_{i=1,2,3} V_{i3}^* V_{i3}^{\phantom{*}}
    \big(C_{\phi q,ii}^{(3)} + C_{\phi q,ii}^{(1)}\big) \, \bar b_L \gamma^\mu b_L \Big] \,.
\end{split}
\label{eq:diagZ}
\end{equation}
This Lagrangian induces a shift of the SM $Z$-boson couplings to
third-generation quarks. We follow the conventions in
Ref.~\cite{Ciuchini:2013pca} and parametrize this shift by
\begin{equation}\label{eq:diagZeff}
\begin{split}
    \delta g_L^t &= \frac{v^2}{2\Lambda^2} \big(C_{\phi q,33}^{(3)} - C_{\phi q,33}^{(1)}\big)\,, \qquad\quad
    \delta g_R^t = - \frac{v^2}{2\Lambda^2} C_{\phi u,33} \,, \\
    \delta g_L^b &= - \frac{v^2}{2\Lambda^2} \sum_{i=1,2,3} V_{i3}^* V_{i3}^{\phantom{*}}
    \big(C_{\phi q,ii}^{(3)} + C_{\phi q,ii}^{(1)}\big) \,.
\end{split}
\end{equation}
The operators $Q_{\phi q,33}^{(1)}$ and $Q_{\phi u,33}$
involve only neutral quark currents, whereas $Q_{\phi q,33}^{(3)}$
will necessarily induce deviations also in the effective $W$
couplings. These charged-current contributions include
\begin{equation}\label{eq:lagFC}
 {\cal L}^\text{}_W = \frac{e}{\sqrt{2} s_w} C_{\phi q,33}^{(3)} \, \sum_i \, V_{3i} \,
 W^+_\mu \, \bar t_L \gamma^\mu d_{L,i} + \text{h.c.}\,. 
\end{equation}
The FCNC contributions relevant for us involve only down-type quarks
and are given by
\begin{equation}
\begin{split}
  {\cal L}^\text{FCNC}_Z = - \frac{e}{2 s_w c_w} \big( C_{\phi
    q,33}^{(3)} + C_{\phi q,33}^{(1)} \big) \sum_{i\neq j}
  V^{\ast}_{3i} V^{\phantom{\ast}}_{3j} \, Z_\mu \, \bar d_{L,i}
  \gamma^\mu d_{L,j} \,.
\label{eq:lagFCNC}
\end{split}
\end{equation}
Note that Eq.~\eqref{eq:lagFCNC} also receives contributions from
the operators in Eq.~\eqref{eq:basis_unbroken_NLO}, but these contributions 
cancel via the unitarity of the CKM matrix (see Sec.~\ref{sec:raredecay})
and we do not include them here. Finally, we remark that
the operators in Eq.~\eqref{eq:basis_unbroken} do not induce
modifications of couplings involving the Higgs boson.

Constraints from flavor physics and EWPO suggest the absence of both
tree-level down-type FCNC transitions and modifications of the $b\bar
b Z$ couplings if NP enters at the TeV scale. Thus, we take $C_{\phi
  q,33}^{(3)} + C_{\phi q,33}^{(1)} = 0$ at the scale $\Lambda$. This
relation can be realized in models with additional vector-like
quarks~\cite{delAguila:2000rc}. Moreover, these models do not generate
four-fermion operators at tree level.
Under these assumptions, rare meson decays and the effective bbZ
couplings are sensitive to anomalous ttZ couplings generated by the
operators in Eq.~\eqref{eq:basis_unbroken}.

\section{Phenomenology\label{sec:pheno}}

\subsection{t-channel single top-quark production\label{sec:singlet}}

The operator $Q_{\phi q,33}^{(3)}$ induces a tree-level correction to
the $Wtb$ coupling given by Eq.~\eqref{eq:lagFC}. Both the ATLAS and
the CMS collaborations have recently measured the t-channel single
top-quark production cross section~\cite{ATLAS-CONF-2014-007,
  Khachatryan:2014iya}, which constitutes a direct measurement of the
$Wtb$ coupling.  In particular, they report the measured cross-section
normalized with respect to the SM expectation,
$\sqrt{\sigma_{\text{t-ch}}(t) /
  \sigma_{\text{t-ch}}^{\text{theo}}(t)}\equiv R_\sigma$.  The
measured values reported by the two collaborations are
\begin{equation}
	R_\sigma = \left\{
	\begin{aligned} &0.97\phantom{0} \pm 0.10\quad\,\text{  [ATLAS]}\\
	 & 0.998 \pm 0.041 \quad\text{[CMS]}
	\end{aligned}\right.
\end{equation}
where we have combined the experimental and theoretical errors in
quadrature.
In our analysis this ratio is given by
\begin{equation}
R_\sigma = 1+v^2 C_{\phi q,33}^{(3)}/\Lambda^2\,.
\label{}
\end{equation}
Combining the measurements, we find
\begin{equation}\label{eq:boundRsigma}
v^2 C_{\phi q,33}^{(3)}/\Lambda^2=-0.006\pm0.038\,.
\end{equation}

\subsection{Electroweak precision observables}\label{sec:EWPO}

\begin{figure}[!t]
\begin{center}
\includegraphics[width=0.3\textwidth]{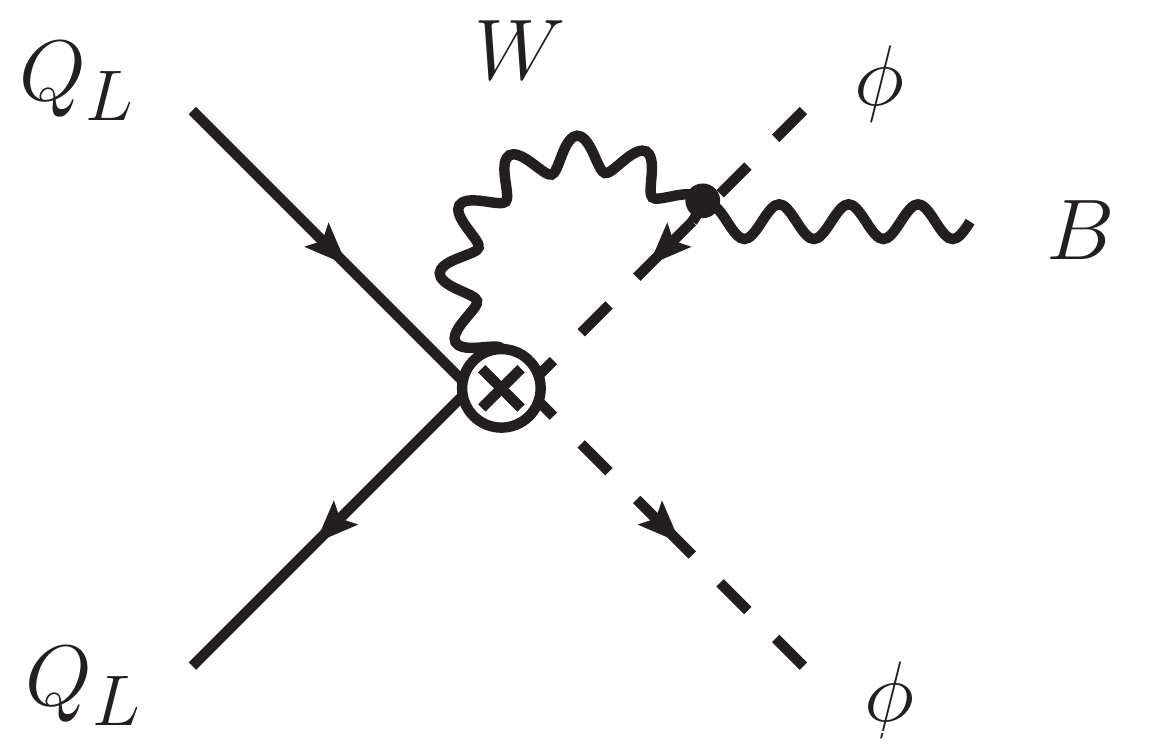}~~~~~~
\includegraphics[width=0.25\textwidth]{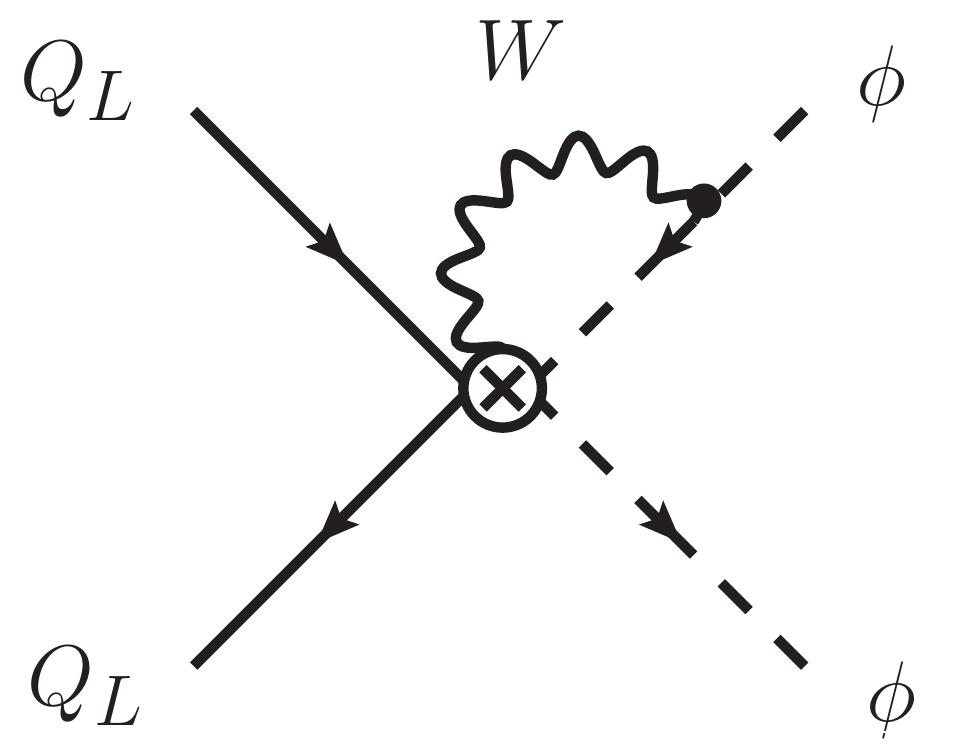}
\end{center}
\caption{Diagrams that induce non-zero mixing proportional to gauge
  couplings of $Q_{\phi q,33}^{(3)}$ into itself. 
  All other diagrams cancel among themselves or against
  contributions from the field renormalization. Note that there are
  additional contributions from penguin-type insertions. As a check,
  we have calculated the mixing both for external states including a
  $B$ gauge boson (left) and for external states not including a $B$
  gauge boson (right) and find the same result.\label{fig:mix}  }
\end{figure}

We saw in Sec.~\ref{sec:tbph} that it is phenomenologically 
reasonable to assume $C_{\phi q,33}^{(3)}+C_{\phi q,33}^{(1)}=0$ at the scale~$\Lambda$.
In particular, this implies $\delta g_{L}^{b}(\Lambda)=0$. Operator
mixing will reintroduce a non-zero $\delta g_{L}^{b}$ at the
electroweak scale. The relevant parts of the RGE are
\begin{equation}\label{eq:RGE}
\begin{split}
 \mu \frac{d}{d\mu} C_{\phi q,33}^{(3)} &= \frac{y_t^2}{16\pi^2}
 \left(8C_{\phi q,33}^{(3)} - 3C_{\phi q,33}^{(1)}\right) -
 \frac{g_2^2}{16\pi^2} \frac{11}{3} C_{\phi q,33}^{(3)}\,, \\
 \mu \frac{d}{d\mu} C_{\phi q,33}^{(1)} &= \frac{y_t^2}{16\pi^2}
 \left(-9C_{\phi q,33}^{(3)}+10C_{\phi q,33}^{(1)}-C_{\phi u}\right) 
 + \frac{g_1^2}{16\pi^2} \frac{1}{9}\big(5C_{\phi q,33}^{(1)}+4C_{\phi u}\big) \,,\\
 \mu \frac{d}{d\mu} C_{\phi q,11}^{(3)} &=\mu \frac{d}{d\mu} C_{\phi q,22}^{(3)} =
 \frac{g_2^2}{16\pi^2}2C_{\phi q,33}^{(3)} \,,\\ 
 \mu \frac{d}{d\mu} C_{\phi q,11}^{(1)} &=\mu \frac{d}{d\mu} C_{\phi q,22}^{(1)} =
 \frac{g_1^2}{16\pi^2}\frac{2}{9}(C_{\phi q,33}^{(1)}+2C_{\phi u}) \,,
\end{split}
\end{equation}
with $y_t$, $g_1$, and $g_2$ the SM top-Yukawa coupling and the
electroweak gauge couplings in the conventions of
Ref.~\cite{Denner:1991kt}, respectively.  The terms proportional to
the top-quark Yukawa coupling have been presented in
Ref.~\cite{Jenkins:2013wua} and those proportional to the gauge
couplings in Ref.~\cite{Alonso:2013hga}. However, our result for the
latter differs from the one of Ref.~\cite{Alonso:2013hga} in the last
term of the first line in Eq.~\eqref{eq:RGE}. We have calculated the
mixing in a general $R_\xi$ gauge and find a gauge-independent result.
While we agree with Ref.~\cite{Alonso:2013hga} on the terms arising
from penguin-type insertions of the operators, we find an additional
contribution for the mixing of $Q_{\phi q,33}^{(3)}$ into itself. It
arises from the diagrams shown in Figure~\ref{fig:mix}. An additional
non-trivial check of our result can be found in the next
section\footnote{The authors of Ref.~\cite{Alonso:2013hga} confirmed
  that this contribution was accidentally omitted in their result;
  however, a corresponding term appears in their anomalous dimension
  of $Q_{\phi l}^{(3)}$.}.

Solving the RGE in Eqs.~\eqref{eq:RGE}, we determine
the Wilson coefficients at the electroweak scale:
\begin{equation}\label{eq:RGEst}
\begin{split}
  C_{\phi q,33}^{(3)}(\mu_W) & = C_{\phi q,33}^{(3)}(\Lambda) + \bigg\{
  \frac{y_t^2}{16\pi^2} \bigg[ 8 C_{\phi q,33}^{(3)}(\Lambda) - 3
    C_{\phi q,33}^{(1)}(\Lambda) \bigg] - \frac{g_2^2}{16\pi^2} \frac{11}{3} C_{\phi q,33}^{(3)}(\Lambda) 
  \bigg\} \log \frac{\mu_W}{\Lambda} \,, \\[3mm]
  C_{\phi q,33}^{(1)}(\mu_W) & = C_{\phi q,33}^{(1)}(\Lambda) + \bigg\{ \frac{y_t^2}{16\pi^2} \bigg[ 
    10C_{\phi q,33}^{(1)}(\Lambda) -9C_{\phi q,33}^{(3)}(\Lambda) - C_{\phi u}(\Lambda)
    \bigg] \\ & \hspace*{7em} + \frac{g_1^2}{16\pi^2}
  \frac{1}{9}\bigg[5C_{\phi q,33}^{(1)}(\Lambda)+4C_{\phi
      u}(\Lambda)\bigg] \bigg\} \log \frac{\mu_W}{\Lambda}\,, \\[3mm] 
  C_{\phi q,11}^{(3)}(\mu_W) & = C_{\phi q,22}^{(3)}(\mu_W) =
  \frac{g_2^2}{8\pi^2} C_{\phi q,33}^{(3)}(\Lambda) 
   \log \frac{\mu_W}{\Lambda}\,, \\[3mm]
  C_{\phi q,11}^{(1)}(\mu_W) & = C_{\phi q,22}^{(1)}(\mu_W) = \frac{g_1^2}{16\pi^2}
  \frac{2}{9}\bigg[C_{\phi q,33}^{(1)}(\Lambda)+2C_{\phi
      u}(\Lambda)\bigg] \log \frac{\mu_W}{\Lambda}\,.
\end{split}
\end{equation}
We used the leading-log approximation to relate the Wilson
coefficient at the scale $\Lambda$ to the electroweak scale
$\mu_W$. Using Eq.~\eqref{eq:diagZeff}, we find
\begin{equation}\label{eq:deltagLb}
\begin{split}
 \delta g_L^b & = - \frac{v^2}{2\Lambda^2}
 \frac{\alpha}{4\pi} \bigg\{ V_{33}^* V_{33}^{\phantom{*}} \bigg[
 \frac{x_t}{2 s_w^2} \Big( 
    8 C_{\phi q,33}^{(1)}(\Lambda) - C_{\phi u}(\Lambda)
    \Big) + \frac{17c_w^2 + s_w^2}{3s_w^2 c_w^2} 
    C_{\phi q,33}^{(1)}(\Lambda) \bigg] \\[3mm] & \hspace{3.25cm} + \bigg[
  \frac{2s_w^2-18c_w^2}{9s_w^2c_w^2} C_{\phi q,33}^{(1)}(\Lambda) +
  \frac{4}{9c_w^2} C_{\phi u}(\Lambda) \bigg] \bigg\} \log \frac{\mu_W}{\Lambda} \,.
\end{split}
\end{equation}
Here, we defined $x_t \equiv m_t^2/M_W^2$ and used the relation
$C_{\phi q,33}^{(3)}(\Lambda) + C_{\phi q,33}^{(1)}(\Lambda) =0$.

The above corrections lead to a shift in the parameter $\epsilon_b$,
defined in Ref.~\cite{Altarelli:1991fk}, given by $\delta \epsilon_b =
\delta g_L^b$. The authors of Ref.~\cite{Rontsch:2014cca} have
compared their direct constraints on anomalous $\ttZ$ couplings with
the indirect constraint derived from $\delta \epsilon_b$. They use the
expression for $\delta \epsilon_b$ given in Ref.~\cite{Larios:1999au}
while using the same effective operators as in our work.  However, the
calculation in Ref.~\cite{Larios:1999au} has been carried out in a
different framework, namely, the non-linearly realized electroweak
chiral Lagrangian~\cite{Larios:1997dc}. Their result therefore does
not agree with ours, and the expression for $\delta \epsilon_b$ in
Ref.~\cite{Rontsch:2014cca} should be replaced by the one in
Eq.~\eqref{eq:deltagLb}.

Quantum corrections also induce the mixing of the operators in
Eq.~\eqref{eq:basis_unbroken} into $Q_{\phi D}$ in
Eq.~\eqref{eq:basis_unbroken_NLO}. This operator is tightly
constrained by EWPO, since it leads to the universal oblique $T$
parameter~\cite{Peskin:1990zt, Peskin:1991sw, Grinstein:1991cd}. The
mixing is given by~\cite{Jenkins:2013wua,Alonso:2013hga}
\begin{equation}
	16\pi^2\mu\frac{d}{d\mu}\,C_{\phi D} = \frac{8}{3}g_1^2\left(C_{\phi
          q,33}^{(1)} + 2C_{\phi u,33}\right) + 24 y_t^2\left(C_{\phi
          q,33}^{(1)} - C_{\phi u,33}\right) \,. 
\end{equation}
Accordingly, we obtain the following expression for the $T$ parameter
(cf. Ref.~\cite{Skiba:2010xn})
\begin{equation}\label{eq:DeltaT}
\begin{split}
	T &= - \frac{v^2}{2\alpha \Lambda^2} C_{\phi D}\\
	&=  -\frac{v^2}{\Lambda^2} \left[\frac{1}{3\pi c_w^2} \left(C_{\phi
            q,33}^{(1)} + 2 C_{\phi u,33}\right) + \frac{3x_t}{2\pi s_w^2} \left(C_{\phi
            q,33}^{(1)} - C_{\phi u,33}\right) \right]\log\frac{\mu_W}{\Lambda} \,.
\end{split}
\end{equation}
We checked that the $T$ parameter obtained via the above RGE analysis
agrees with a direct computation of the vacuum-polarization
diagrams. 

The $T$ parameter is directly related to the quantity $\epsilon_1$,
defined in Ref.~\cite{Altarelli:1993sz}, via $\epsilon_1 \equiv \alpha
T$. The term proportional to $x_t$ in Eq.~\eqref{eq:DeltaT} can be
deduced from the corresponding result for $\delta \epsilon_1$ in
Ref.~\cite{Malkawi:1994tg}; we agree with that result. However, our
result disagrees with that quoted in Ref.~\cite{Rontsch:2014cca},
where the contributions from the modified charged current have not
been included.

Note that there are no contributions to the $S$ parameter within our
setup because the operators in Eq.\eqref{eq:basis_unbroken} do not mix
into $Q_{\phi W B} \equiv (\phi^\dagger \sigma^a \phi) W_{\mu\nu}^a
B^{\mu\nu}$. Vertex corrections could, however, lead to an indirect
contribution to $S$ (see e.g. Ref.~\cite{Gupta:2014rxa}). This is
illustrated, for instance, by the correlation of $\delta \epsilon_3$
and $\delta \epsilon_b$~\cite{Ciuchini:2014dea}. We neglect these
effects for simplicity. On the other hand, the operators in
Eq.\eqref{eq:basis_unbroken} mix into $Q_{\phi l,11}^{(3)} \equiv
(\phi^\dagger i\!  \stackrel{\leftrightarrow}{D_\mu^a} \phi) (\bar
L_{L,1} \gamma^\mu \sigma^a L_{L,1})$ according to
\begin{equation}\label{eq:emix}
 \mu \frac{d}{d\mu} C_{\phi l,11}^{(3)} = 2 \frac{g_2^2}{16\pi^2}
 C_{\phi q,33}^{(3)}\,.
\end{equation}
This will induce a shift in the $Z$ coupling to electrons, defined by 
\begin{equation}
    \delta g_L^e = - \frac{v^2}{2\Lambda^2} \big(C_{\phi l,11}^{(3)} +
    C_{\phi l,11}^{(1)}\big) \,.
\end{equation}
Solving Eq.~\eqref{eq:emix} we find
\begin{equation}
 \delta g_L^e = - \frac{v^2}{\Lambda^2} \frac{\alpha}{4\pi s_w^2}
 C_{\phi q,33}^{(3)}(\Lambda)\log \frac{\mu_W}{\Lambda} \,.
\end{equation}
Comparing with LEP data~\cite{ALEPH:2005ab} we see that the resulting
bound on $C_{\phi q,33}^{(3)}$ is weaker than that from $\delta g_L^b$
and we do not consider this constraint further. Similar considerations
apply also for the corresponding $\mu$ and $\tau$ couplings. A more
comprehensive analysis taking all of these effects into account might
be worthwhile.

\subsection{Rare meson decays}\label{sec:raredecay}

We now advocate the use of rare meson decays to constrain anomalous
$\ttZ$ couplings. Recall that in this work we assume $C_{\phi
  q,33}^{(3)} + C_{\phi q,33}^{(1)} = 0$, and thus the absence of
tree-level FCNC transitions at the scale $\Lambda$.  Operator mixing
reintroduces these transitions at lower scales. We calculate the
running of the Wilson coefficients from $\Lambda$ to the electroweak
scale, where we match onto the five-flavor effective theory. We then
compute the modifications of the rare meson decay rates, which allows
us to bound the Wilson coefficients at the scale $\Lambda$.  We focus
here on the processes $B_{(s,d)} \to \mu^+ \mu^-$ and $K\to \pi \nu
\bar \nu$. The reason is that all these decays are dominated by the
$Z$-penguin within the SM and are thus the best candidates to
constrain anomalous $\ttZ$ couplings.

The part of the RGE relevant for the rare meson decays is the same as
in Eq.~\eqref{eq:RGE}, with the addition
\begin{equation}\label{eq:RGE4f}
\begin{split}
 \mu \frac{d}{d\mu} C_{lq}^{(3)} = - \frac{g_2^2}{16\pi^2} \frac{1}{3} C_{\phi q,33}^{(3)} \,,\qquad
 \mu \frac{d}{d\mu} C_{lq}^{(1)} = \frac{g_1^2}{16\pi^2} \frac{1}{3} C_{\phi q,33}^{(1)} \,.
\end{split}
\end{equation}
As a non-trivial check of our calculation we reproduce the logarithmic
part of the loop functions $f_{\nu\bar\nu}^{(j)}$ given in
Ref.~\cite{Drobnak:2011aa}. Solving the equations~\eqref{eq:RGE4f}, we
find the following expressions for the Wilson coefficients at the
electroweak scale:
\begin{equation}
\begin{split}
  C_{lq}^{(3)}(\mu_W) & = C_{lq}^{(3)}(\Lambda) + \frac{1}{3} C_{\phi q,33}^{(3)}(\Lambda)
  \frac{g_2^2}{16\pi^2} \log \frac{\mu_W}{\Lambda} \,, \\[3mm]
  C_{lq}^{(1)}(\mu_W) & = C_{lq}^{(1)}(\Lambda) - \frac{1}{3}
  C_{\phi q,33}^{(1)}(\Lambda)\frac{g_1^2}{16\pi^2}  \log \frac{\mu_W}{\Lambda}\,. 
\end{split}
\end{equation}
The other relevant Wilson coefficients are given in
Eq.~\eqref{eq:RGEst}. Note, however, that now some of the
contributions to the FCNC transitions in Eq.~\eqref{eq:RGEst} cancel
because of the unitarity of the CKM matrix. In particular, $C_{\phi
  q,ii}^{(1)}$ and $C_{\phi q,ii}^{(3)}$ for $i = 1,2$ do not appear
in the final expressions.

The rate of the rare decay $\Bsmumu$ was measured by the
CMS~\cite{Chatrchyan:2013bka} and LHCb~\cite{Aaij:2013aka}
collaborations at the LHC and it was found to be consistent with the 
SM expectations~\cite{Bobeth:2013uxa}. 
The experiments are expected to
reach the sensitivity of observing the SM $\Bdmumu$ rate in the next
LHC run. Within the SM and the restricted set of operators we consider, 
a single operator $Q_A\equiv (\bar b\gamma_\mu\gamma_5
q)(\bar\mu\gamma_5\mu)$ with $q=d,s$ mediates the decays to a very
good approximation. The rates depend on a single hadronic
quantity, the decay constant $f_{B_q}$ of the meson. The
average time-integrated branching ratios then read
\cite{Bobeth:2013uxa}:
\begin{equation} \label{eq:brbmm}
\overline{\mathcal B}_{q\mu} \;=\; \frac{|N|^2 M_{B_q}^3
f_{B_q}^2}{8\pi\,\Gamma^q_H}\, 
\beta_{q\mu}\, r_{q\mu}^2\, |C_A(\mu_b)|^2 \,+\, {\mathcal O}(\alpha_{em})\,, 
\end{equation}
with $M_{B_q}$ the mass of the $B$ meson, $r_{q\mu} = 2
m_\mu/M_{B_q}$, $\beta_{q\mu} = \sqrt{1-r_{q\mu}^2}$, $N =V_{tb}^*
V_{tq} \GF^2 M_W^2/\pi^2$ and $\Gamma^q_H$ the total width of the
heavy mass eigenstate in $B_q$-$\overline{B}_q$ mixing. In general,
the time-integrated rates depend on the details of
$B_q$-$\overline{B}_q$ mixing \cite{DeBruyn:2012wk}, but
Eq.~\eqref{eq:brbmm} holds to a very good approximation for both the
$B_s$ and the $B_d$ system within the SM.  $C_A(\mu_b)$ is the Wilson
coefficient of $Q_A$. It incorporates effects from heavy NP to the SM
contribution and is evolved to the scale of the $B$-meson, $\mu_b\sim
m_b$. The SM part includes perturbative next-to-leading-order (NLO)
QCD \cite{Buchalla:1992zm,Buchalla:1993bv,Misiak:1999yg} as well as
recently calculated next-to-next-to-leading (NNLO) QCD
\cite{Hermann:2013kca} and NLO electroweak \cite{Bobeth:2013tba}
corrections, which we all include in our analysis. However, at leading
order, we have that $C_A = -2 (Y_0(x_t) + \delta Y^{\rm NP})$ where
$Y_0(x_t)$ is the SM loop function and $\delta Y^{\rm NP}$ is the
additional contribution from NP.

Similarly, the branching ratio for $K^+ \to \pi^+ \nu \bar\nu$ can be
written as~\cite{Buchalla:1993wq, Buchalla:1998ba, Isidori:2005xm}
\begin{multline}\label{eq:BR}
  \text{Br} \left(K^+\to\pi^+\nu\bar{\nu}(\gamma)\right) \\ =\kappa_+
  (1+\Delta_{\text{EM}})
  \Bigg[\left(\frac{\text{Im}\lambda_t}{\lambda^5} X_t\right)^2 +
  \left(\frac{\text{Re}\lambda_c}{\lambda} \left(P_c + \delta P_{c,u}
    \right) + \frac{\text{Re}\lambda_t}{\lambda^5} X_t\right)^2
  \Bigg]\,,
\end{multline}
where $\lambda \equiv |V_{us}|$ and $\lambda_i \equiv V_{is}^*
V_{id}^{\phantom{*}}$. The leading contribution is contained in the
function $X_t$ which comprises the top-quark loops. It is known to
NLO in QCD and electroweak interactions~\cite{Buchalla:1998ba, Buchalla:1992zm, Misiak:1999yg,
  Brod:2010hi}. The parameter $P_c$ describes the short-distance
contribution of the charm quark and has been calculated including
NNLO QCD~\cite{Gorbahn:2004my, Buras:2005gr,Buras:2006gb} and NLO electroweak
corrections~\cite{Brod:2008ss}. Other long-distance contributions are
contained in the parameter $\delta P_{c,u}$~\cite{Isidori:2005xm}. The
hadronic matrix element of the low-energy operator is parameterized by
$\kappa_+$. This parameter is extracted from $K_{\ell 3}$ data and
known precisely, including long-distance QED radiative corrections
($\Delta_{\text{EM}}$), and NLO and partially NNLO corrections in
chiral perturbation theory~\cite{Mescia:2007kn, Bijnens:2007xa}.

The branching ratio of the $CP$-violating neutral mode involves only
the top-quark contribution and can be written as\footnote{In fact, in
  our numerics we take into account also the (small) contribution from
  indirect $CP$ violation, proportional to $\epsilon_K$ -- see
  Ref.~\cite{Buchalla:1996fp}.}
\begin{equation}
  \text{Br}\left(K_L\to\pi^0\nu\bar\nu\right)=\kappa_L\left(
    \frac{\text{Im}\lambda_t}{\lambda^5}X_t \right)^2\, .
  \label{eq:brkL}
\end{equation}
Again, the hadronic matrix element can be extracted from the $K_{l3}$
decays, parametrized here by $\kappa_L$~\cite{Mescia:2007kn}.

As in the $\Bqmumu$ decays, we include all NP effects 
as additional contributions to the SM top-quark function 
$X_t = X_t^\text{SM} + \delta X^{\rm NP}$.
At the order we consider, all NP effects originate from the modifications of the 
$\ttZ$ coupling. Thus, these are the same for $\Bqmumu$ and $K\to\pi\nu\bar\nu$
decays
\begin{equation}
\begin{split}
  \delta Y^{\rm NP} = \delta X^{\rm NP} = \frac{x_t}{8} \bigg( C_{\phi u} 
    - \frac{12 + 8 x_t}{x_t} C_{\phi q,33}^{(1)} \bigg)
    \frac{v^2}{\Lambda^2} \log \frac{\mu_W}{\Lambda} \,,  
\end{split}
\end{equation}
where $x_t = m_t^2/M_W^2$ and we used again the relation $C_{\phi
  q,33}^{(3)} = - C_{\phi q,33}^{(1)}$ at the scale~$\Lambda$.

We conclude this section by comparing our work to existing results in
the literature. Rare meson decays have been used to constrain FCNC
$Zqt$ couplings, where $q=u,c$, in Ref.~\cite{Li:2011af, Gong:2013sh},
and to constrain anomalous $Wtb$ couplings in
Ref.~\cite{Drobnak:2011aa}. The corrections to the Wilson coefficients
presented in the above publications amount to one-loop threshold
corrections at the electroweak scale. These corrections are scheme
dependent~\cite{Buras:1991jm, Ciuchini:1993vr}; the scheme dependence
would cancel only when performing the two-loop running of the Wilson
coefficients in the effective theory above the electroweak scale. On
the other hand, the logarithmic dependence of the corrections on the
scale $\Lambda$ is scheme independent (since the leading-order
anomalous dimensions are scheme independent). By effectively choosing
a questionably low matching scale, of the order of the $W$-boson mass,
these terms are rendered numerically insignificant in the above
articles.

\section{Numerics and discussion\label{sec:discussion}}

In this section we present the constraints on anomalous $\ttZ$
couplings derived from the observables discussed above. We show the
individual constraints and perform a combined fit using current
experimental data. In addition, we show the impact of future precision
measurements of the rare $B$ and $K$ decay branching ratios.

\begin{table}
\begin{center}
\begin{tabular}{lcr}
  {\bf Observable} & {\bf Value} & {\bf Ref.}\\\hline \hline 
  $T$ & $0.08\phantom{0} \pm 0.07\phantom{0}$& \cite{Ciuchini:2013pca}\\
  $\delta g_L^{b}$ & $0.0016 \pm 0.0015$ & \cite{Ciuchini:2013pca}\\
  Br($B_s \to \mu^+ \mu^-$) [CMS] & $(3.0^{+1.0}_{-0.9})\times 10^{-9}$ & \cite{Chatrchyan:2013bka}\\[1mm]
  Br($B_s \to \mu^+ \mu^-$) [LHCb] & $(2.9^{+1.1}_{-1.0})\times 10^{-9}$ & \cite{Aaij:2013aka}\\[1mm]
  Br($K^+ \to \pi^+ \nu \bar \nu$) & $(1.73^{+1.15}_{-1.05})\times 10^{-10}$& \cite{Artamonov:2008qb}\\\hline
\end{tabular}
\end{center}
\caption{Numerical input values for our fit.\label{tab:input}}
\end{table}

\begin{figure}[!t]
\hspace*{-1em}\includegraphics[]{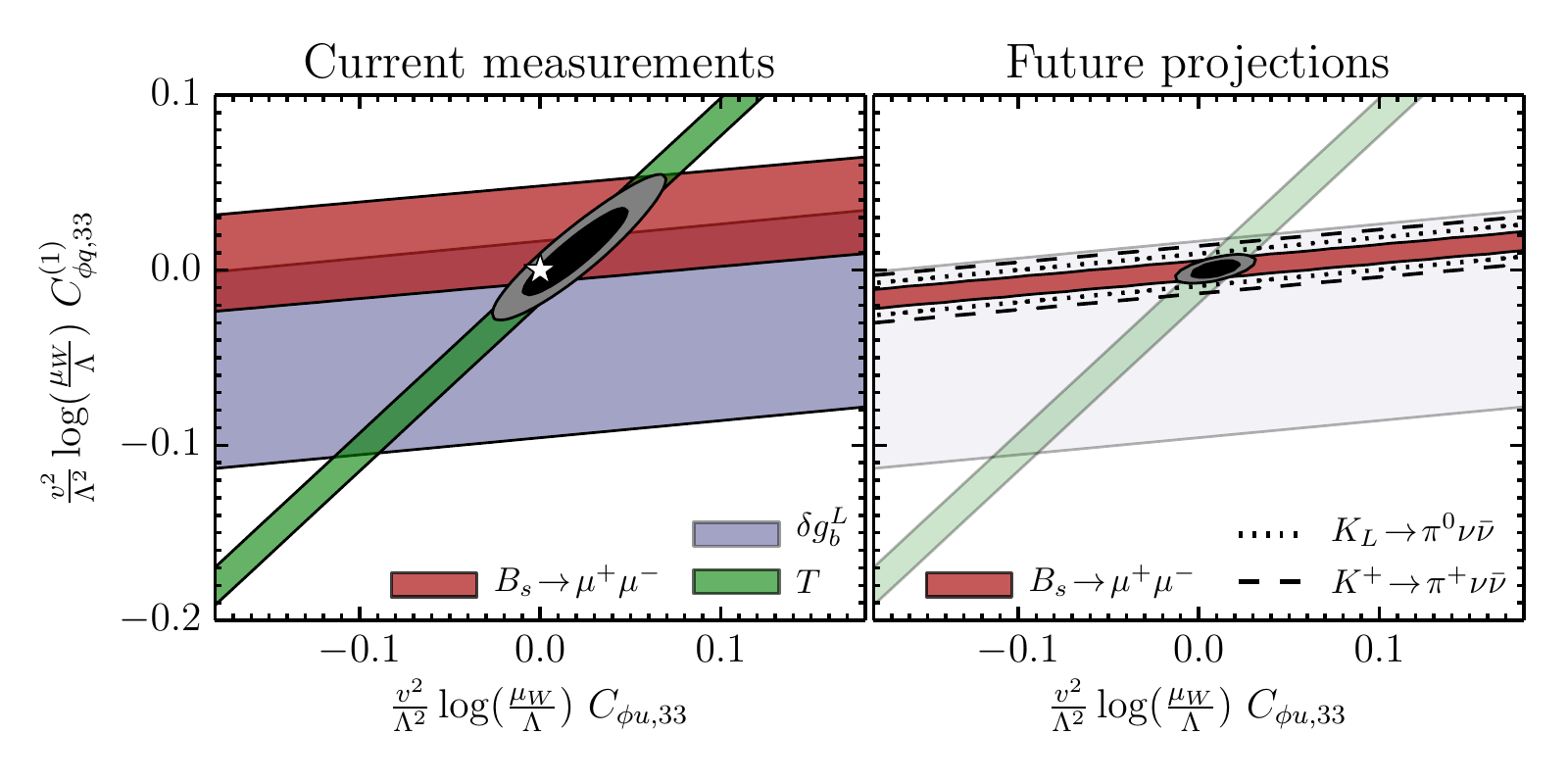} 
\caption{The preferred regions at $68\%$ and $95\%$ CL from our
  combined fit to EWPO and rare decays are shown as the dark-gray and
  light-gray ellipses, respectively. The colored bands show the $68\%$
  CL constraints from the individual observables. The star denotes the
  SM value. 
  \label{fig:present}}
\end{figure}

Our main results are summarized in Fig.~\ref{fig:present}. In the left
panel we show the individual $68\%$ CL
regions resulting from the
measurement of the $T$ parameter, $\delta g_L^b$,
Br($K^+\to\pi^+\nu\bar\nu$), and Br($B_s\to\mu^+\mu^-$). In addition,
we show the region compatible with the measurements in
Table~\ref{tab:input} at $68\%$ and $95\%$ CL. 

We find that the branching ratio of $B_s\to\mu^+\mu^-$ and the $T$
parameter currently lead to the most stringent constraints. In
particular, the combination of the two leads to a strong bound on both
Wilson coefficients $C_{\phi q, 33}^{(1)}\log(\mu_W/\Lambda)\, v^2/
\Lambda^2$ and $C_{\phi u, 33}\log(\mu_W/\Lambda)\, v^2/ \Lambda^2$,
of the order of a few percent.
We note that t-channel single-top
production leads to the bound $-0.032 < v^2 C_{\phi q, 33}^{(1)} /
\Lambda^2 < 0.044$ (cf.~Eq.~\eqref{eq:boundRsigma}). This is weaker
than the indirect bounds and leaves $C_{\phi u, 33}$ 
completely unconstrained.

In the future, we expect improvements in the measurement of
Br($B_s\to\mu^+\mu^-$), with a final uncertainty of
$\sim\!5$\%~\cite{Bediaga:2012py}. In addition, various experiments
plan to measure the branching ratios of the rare $K$ decays with high
precision.  The NA62 experiment at CERN aims at a final precision of
$\sim\!10$\% for the charged mode, which could be improved to
$\sim\!3$\% by an experiment at Fermilab~\cite{Kronfeld:2013uoa}. The
KOTO experiment aims at a similar precision for the neutral mode. On
the other hand, the bounds from EWPO are mainly obtained from fits to
LEP data and we do not expect any significant improvement within the
next few years. In the right panel of Fig.~\ref{fig:present} we show
our future projections. As an illustration we assume a branching ratio
measurement of all three rare decay modes with the SM central values
and a precision of 5\%. We keep the current constraints from the EWPO,
but note that these bounds could be improved at future $e^+ e^-$
colliders~\cite{Gomez-Ceballos:2013zzn}.

The indirect constraints on the anomalous $\ttZ$ couplings are 
much stronger than the constraints from direct searches,
i.e.~from $t\bar t + Z$ production, even after a high-luminosity
upgrade of the LHC. For instance, the authors of
Ref.~\cite{Rontsch:2014cca} give the bounds $-0.04 < v^2/\Lambda^2 \,
C_{\phi q,33}^{(1)} < 0.19$ and $-0.13 < v^2/\Lambda^2 \, C_{\phi
  u,33} < 0.32$, assuming $3000\,\text{fb}^{-1}$ of data.  However,
one has to keep in mind that indirect constraints rely on a set of
assumptions. In this work we assumed i) only $C_{\phi q,33}^{(3)}$,
$C_{\phi q,33}^{(1)}$, and $C_{\phi u,33}$ receive non-zero initial
conditions at the scale $\Lambda$; ii) $C_{\phi q,33}^{(3)} + C_{\phi
  q,33}^{(1)} = 0$ at $\Lambda$; iii) only the top-quark Yukawa coupling is
non-vanishing.

Assumption i) is compatible only with NP models with non-trivial
flavor structure, while assumption ii) can be motivated by explicit
models~\cite{delAguila:2000rc}. A simple way to deviate from
assumption iii) is to consider models with a large enhancement of the
bottom-quark Yukawa coupling; a generic example is a two Higgs-doublet
model with large $\tan\beta$. The large bottom-Yukawa coupling will
induce flavor off-diagonal versions of the operators in
Eq.~\eqref{eq:basis_unbroken} and Eq.~\eqref{eq:basis_unbroken_NLO},
via the MFV counting. These off-diagonal operators lead to additional
contributions to FCNC top decays and $D^0 - \overline{D^0}$ mixing. In
order to relate these observables to $\ttZ$ couplings, we assume
MFV. Thus the resulting constraints on anomalous $\ttZ$ couplings are
suppressed by CKM-matrix elements. As an illustrative example consider
an extreme case where the bottom-Yukawa coupling is much larger than
the top-Yukawa coupling. In this case, we have $C_{\phi q,23}^{(3)}
\sim \lambda^2 C_{\phi q,33}^{(3)}$ etc., where $\lambda \equiv
|V_{us}| \approx 0.22$ is the Wolfenstein parameter. Then $D^0 -
\overline{D^0}$ mixing is suppressed by $\lambda^{10} \approx 10^{-7}$
and thus completely negligible. Also, the branching ratio for $t \to c
Z$ is
\begin{equation}
 	\text{Br}(t\to cZ) \simeq \frac{\lambda^4v^4}{\Lambda^4}
        \left[\left(C_{\phi q,33}^{(3)}-C_{\phi q,33}^{(1)}\right)^2+C_{\phi
          u,33}^2\right]\,.
\end{equation}
Using the present bound Br$(t \to c Z)<0.05\%$ given by the CMS
collaboration~\cite{Chatrchyan:2013nwa} we see that the resulting
bounds are not competitive with bounds from EWPO and rare $B$/$K$
decays.

Note that the off-diagonal operators will also lead to additional
contributions to rare $B$/$K$ decays and anomalous $\bbZ$ couplings.
The generalization of our assumption ii) can be used to eliminate such
contribution from these off-diagonal operators~\cite{delAguila:2000rc}.

More generally, all rare decays in the down sector which receive a
$Z$-penguin contribution can be used to obtain bounds on anomalous
$\ttZ$ couplings with our method. Suitable decays which will be
measured in the future include $\Bdmumu$~\cite{Hewett:2012ns} and $B
\to K \nu \bar \nu$~\cite{Aushev:2010bq}. It would be interesting to
allow for complex Wilson coefficients of the operators in
Eq.~\eqref{eq:basis_unbroken} and study their effect on $CP$ violation
in rare meson decays.

To conclude, in this work we studied the effects of dimension-six
operators, generating anomalous vector and axial-vector $\ttZ$
couplings at tree-level, on precision observables. In particular, we
advocate the use of rare $K$ and $B$ meson decays to obtain strong
constraints on $\ttZ$ couplings.

\phantomsection
\addcontentsline{toc}{section}{Acknowledgements}
\section*{Acknowledgements}

We thank Jernej Kamenik and Jure Zupan for useful comments on the
manuscript, Yosef Nir and Gilad Perez for insightful discussions, and
Marco Ciuchini, Alex Pomarol and Luca Silvestrini for helpful
correspondence concerning the $S$ parameter. This work was supported
in part by National Science Foundation Grant No. PHYS-1066293 and the
hospitality of the Aspen Center for Physics. J.B. thanks the Perimeter
Institute for Theoretical Physics for their hospitality. Research at
Perimeter Institute is supported by the Government of Canada through
Industry Canada and by the Province of Ontario through the Ministry of
Economic Development \& Innovation. The work of J.B. is supported in
part by the U.S. National Science Foundation under CAREER Grant
PHY-1151392.  The work of P.U. is supported in part by the DOE grant
DE- SC0011784.  A.G.\ acknowledges the support by the Slovenian
Research Agency (ARRS).


\cleardoublepage
\phantomsection \addcontentsline{toc}{section}{References}
\bibliography{paper} \bibliographystyle{kpinunu_Xt}


\end{document}